\documentclass[preprint,12pt]{elsarticle}
\usepackage{amssymb,epsfig}
\journal{Physica A}

\newcommand{\mean}[1]{\left\langle#1\right\rangle}
\newcommand{\rysunek}[3]{\begin{figure}[ht] \centerline{\epsfig{file=#2,width=0.8\columnwidth}} \caption{#3} \label{#1} \end{figure}}

\begin{document}
\begin{frontmatter}



\title{Hierarchy depth in directed networks}


\author{Krzysztof Suchecki}
\author{Janusz Ho\l yst}

\address{Faculty of Physics and Center of Excellence for Complex Systems Research\\Warsaw University of Technology\\Koszykowa 75, 00-662 Warszawa, PL}

\begin{abstract}
We explore depth measures for flow hierarchy in directed networks. We define two measures -- rooted depth and relative depth, and discuss differences between them. We investigate how the two measures behave in random Erdos-Renyi graphs of different sizes and densities and explain obtained results.
\end{abstract}


\end{frontmatter}

% \linenumbers
\section{Introduction}
The concept of hierarchy has been used for systems described with networks for a significant time and for different purposes. It has been considered in context of elections \cite{ksjh_shimbel}, investigation of road network geometry \cite{ksjh_okabe} and neural networks \cite{ksjh_willcox}. More recently, the concept of hierarchy has been used for social systems \cite{ksjh_bonabeau,ksjh_sousa,ksjh_stauffer,ksjh_gallos,ksjh_valverde}, artificial neural networks \cite{ksjh_yamashita}, financial markets \cite{ksjh_tumminello} and communication networks \cite{ksjh_wang}. It has also been used on a more theoretical level, to distinguish and explain specific network properties \cite{ksjh_ravasz}. While the concept of hierarchy has been extensively used, it is continuously lacking a definition. There are papers that attempt to remedy that by introducing more formal definition and measures \cite{ksjh_trusina,ksjh_mones,ksjh_murtra}. The paper by Corominas-Murtra \cite{ksjh_murtra} introduces definition of causal 
graphs and perfect hierarchical structure, and introduces measure of hierarchy that is based on similarity of a given graph to a perfect hierarchy. The paper by Mones et. al. \cite{ksjh_mones} distinguishes between three types of hierarchy: order (as in \cite{ksjh_bonabeau,ksjh_sousa,ksjh_stauffer,ksjh_gallos,ksjh_valverde}), nested (as in \cite{ksjh_ravasz}) and flow (as in that paper and \cite{ksjh_murtra}). The order hierarchy is a simple ordering of elements of a set along one dimensional axis. The nested hierarchy is tied to the community structure and multi-scale organization of the network. The flow hierarchy is basically a causal structure, with (using tree analogy) the ``root'' being the origin of all signals or flows, and ``leaves'' being simply receivers. In this paper, we focus on the flow hierarchy. As pointed out in \cite{ksjh_mones} order hierarchy can be seen as simple flow hierarchy, and nested hierarchy can be represented by a flow hierarchy. In our work, we intend not to make another 
definition of hierarchy, but to measure a single aspect of it -- its depth. We introduce two depth measures for directed networks and discuss their behavior and differences between them. We also show how the existence of cycles can be taken into account and how they influence depth measures.

\section{Definitions of depth}
\label{ksjh_sect_definitions}
%The definition of depth in a directed network is not trivial. It is possible to define it in different ways, that due to complexity of the topology may not give same or even similar results. We will consider the depth in the context ot information flows or command chains, where directed nature of edges mean that one vertex can influence another, but not necessarily the other way around. One of the concepts we are interested in, while exploring the depth definitions, is the concept of hierarchy and how to measure or even define it. We will consider and compare two definitions of depth. First one, which we will call \emph{rooted depth} is based on the distance from root, and the second one, which we will call \emph{relative depth} is based on relations of neighbors.\\
We define depth measures for directed networks, that are composed of vertices connected by arcs (directed edges). In general, they are applicable to any directed network, even if it contains cycles. The depth is defined in the context of flow hierarchies, and is therefore most meaningful for networks that have a certain overall direction of all arcs.

The \emph{rooted depth} is a value defined for each vertex in a network, and is the distance of that vertex from a root. Use of this definition requires a fixed root vertex. Let us start simple and first consider a tree, where root is well defined. We are considering directed networks and consider tree where arcs are all directed from root and towards leaves. The root is a vertex with only outgoing arcs, while leaves have only incoming arcs. Each vertex $i$ is at certain depth $d_i$ in the network, which we define as the distance from root to the vertex $l_{ri}$. This means that all vertices are organized into distinct depth levels.\\
Let us now generalize to any graph with a specified root, which we will call \emph{rooted network}. The root is the only vertex that has only outgoing arcs, and no incoming arcs ($k^{in}_r=0$). Leaves are vertices without outgoing arcs, but some incoming arcs. The definition remains same without any changes, the distance being the length of shortest path. Because of the existence of multiple paths of different lengths, it is possible however, that an arc connects higher depth vertex with lower depth vertex (Figure \ref{ksjh_fig_simplehierarchy}). Because the definition uses shortest path, it effectively ignores any arcs that are not part of a shortest path. This definition works for networks with loops and effectively ignores them. It is easy to understand, as at least part of the loop has to be not a shortest path from root to somewhere, so it simply is ignored.\\

\rysunek{ksjh_fig_simplehierarchy}{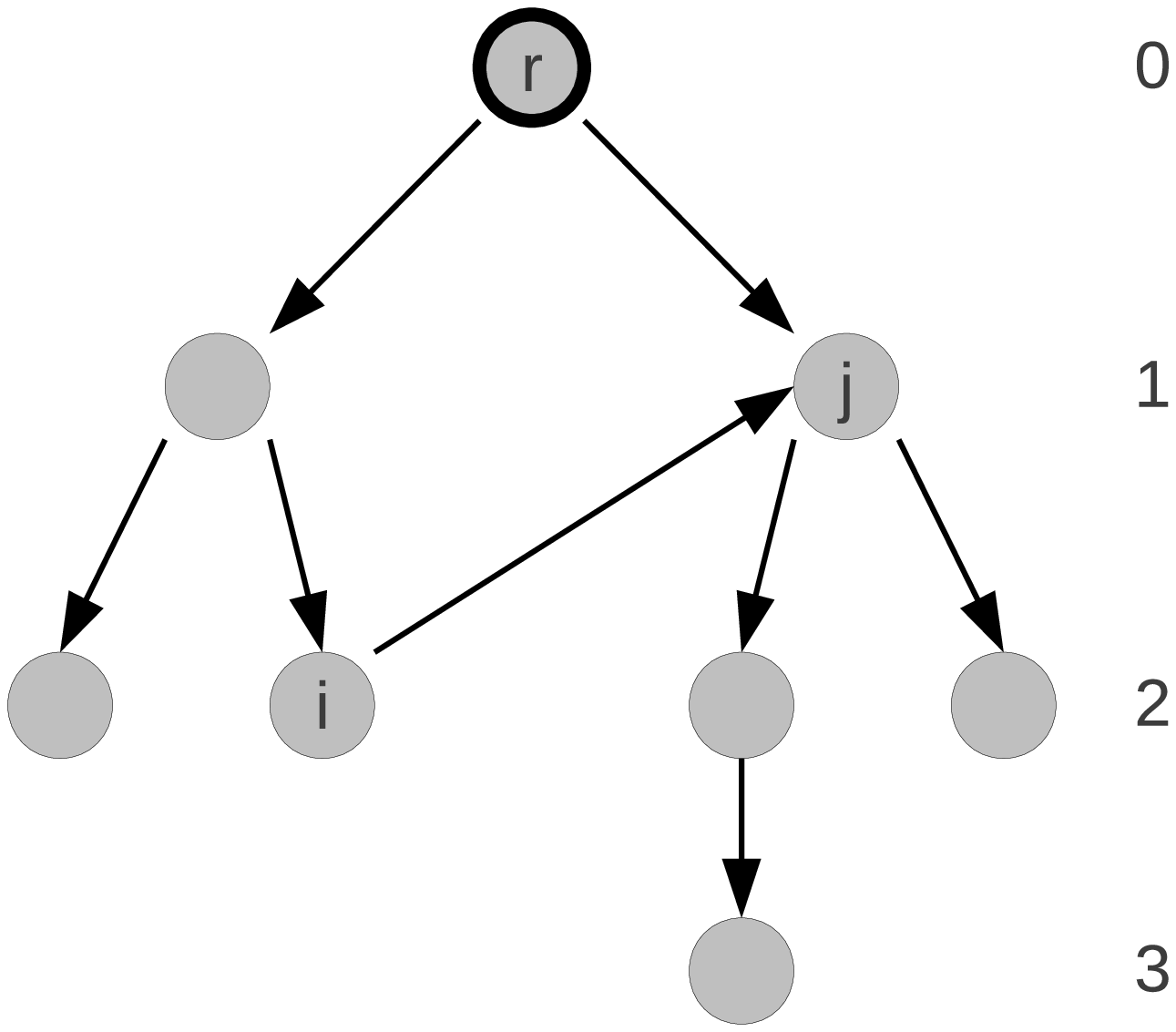}{Rooted network of $3$ depth levels, with root $r$ marked by thick boundary. Vertex $i$ is at depth $2$, while vertex $j$, connected by incoming arc from $i$ is at depth $1$, showing that globally defined hierarchy does not necessarily describe local structure. Average depth (calculated for leaves) is $2 1/3$, because leaves are at different depths. The arc from $i$ to $j$ is effectively ignored when depth levels are calculated. The numbers on the right show depth levels.}

If we define any vertex that has no incoming arcs as a root, we may treat any directed graph as a sum of several rooted networks (Figure \ref{ksjh_fig_acyclic}). Each root vertex has a part of network reachable form it, forming its own \emph{rooted component}, which may encompass whole network, but it not necessarily does. Given multiple roots, a given vertex $i$ has several possibly different depths, depending which root we will look at. Reasonable solution is to average the depths from different roots to obtain a single depth value for a vertex. Additional issue comes from fact, that existence of loops means that no roots may exist in the whole network. In such case, the definition does not work, unless we perform loop collapse (see below).\\

\rysunek{ksjh_fig_acyclic}{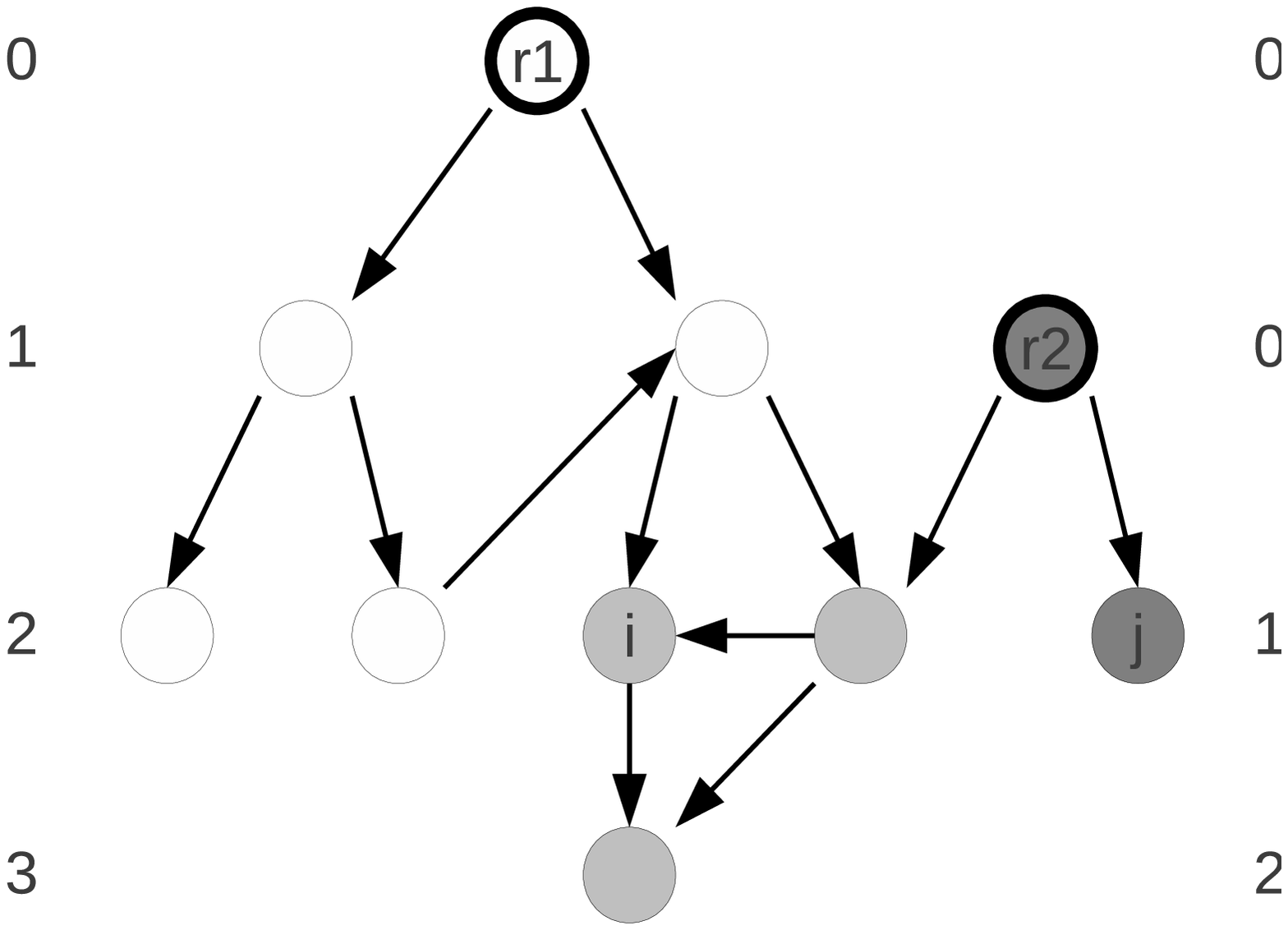}{Acyclic directed graph and its rooted depth. White vertices belong only to rooted component of root $r1$, while dark grey only to component of root $r2$. Light grey vertices belong to both rooted components. Depth levels of vertices (left for $r1$, right for $r2$) may be different relative to each root. Vertex $i$ is at depth $2$ from root $r1$ and at depth $2$ from $r2$ -- it is generally not possible to draw the depth levels consistently for rooted depth. Note that the graph does not need to be acyclic for the rooted depth to be defined.}

The \emph{relative depth} is defined as a value $d_i$ assigned to each vertex $i$, such that any arc $ij$ always connects from a vertex with lower depth $d_i$ to a vertex with higher depth $d_j$ ($d_i<d_j$). This basically means that we treat the arcs as a ``has depth lesser than'' relation on the set of vertices of the graph, and thus define the depth levels for vertices. Such definition is ambiguous, and to give precise values, we add that the difference must be minimal but at least $1$ ($d_j-d_i\geq1$). Depth values assigned according to such definition correspond to difference between vertex depths of $i$ and $j$ being equal to longest directed path between them, provided such path exists (Figure \ref{ksjh_fig_alternative}). In absence of the path, the difference can be any value. Note that since the values are derived from relations, there is no set ``zero level'' and the values can be set relative to an arbitrary reference value. For convenience, it is reasonable to set vertex with 
least depth to $0$, thus recovering non-negative depth levels like in the rooted depth definition. The definition of relative depth sets a strict inequality relations between vertices that have directed paths between them, meaning that they can indirectly interact in some way if the network is treated as a topology of interactions. The definition works only for acyclic graphs. Any directed loop is unresolvable, just like an inequality $a>b>c>a$ does not have any solution.\\

\rysunek{ksjh_fig_alternative}{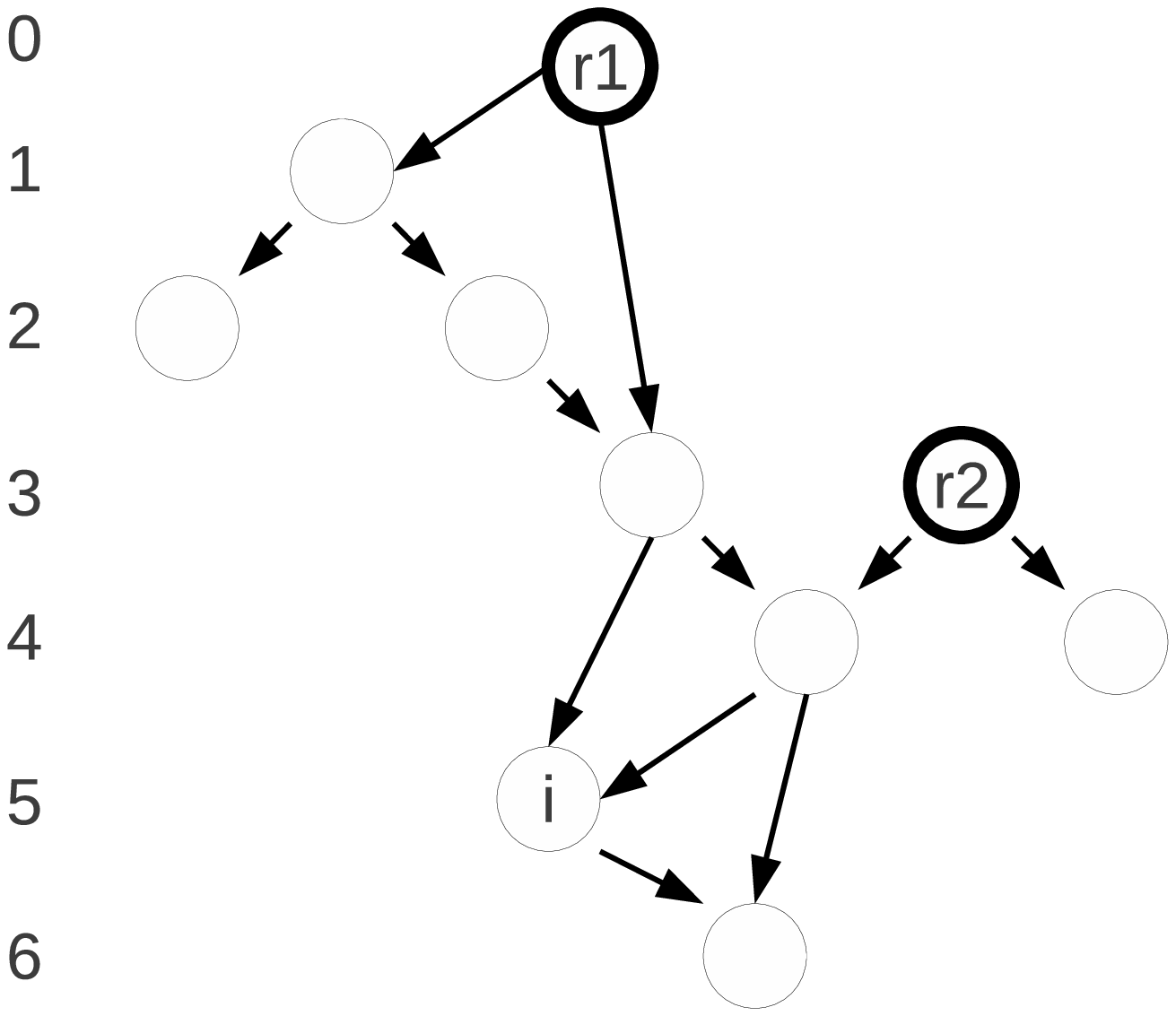}{Acyclic directed graph and its relative depth. The graph is the same as at Figure \ref{ksjh_fig_acyclic}, except the layout being adapted to relative depth values. Note vertex $i$ is at high depth $5$ despite being only 2 arcs away from the depth $0$. Two roots present in the graph, $r1$ and $r2$ are found at depths $0$ and $3$. All arcs always point downwards, by definition of relative depth.}

We can extend the definition to graph containing cycles if we force all vertices belonging to a given cycle to have the same depth value. This means each cycle is at fixed depth level, regardless of the directed arcs it contains. This is however a logical rule -- if all vertices can influence all other vertices in the set, all should be at same depth. Effectively, from the point of view of the definition, a cycle is a single complex vertex. We define \emph{loop collapse} to be a process, where all cycles are substituted with a complex single vertices (see Figure \ref{ksjh_fig_collapse}). Note that the loop collapse is virtual -- the vertices are not really collapsed, they are only treated so for the single purpose of assigning the depth level. All vertices that are component of a collapsed loop are assigned same depth. If all loops are collapsed, then we recover a directed acyclic graph, and relative depth definition can be used to calculate depth levels. Note that it is possible to use loop collapse 
independently from relative depth definition, for example using it alongside rooted depth definition, in which case it changes the depth levels.\\

\rysunek{ksjh_fig_collapse}{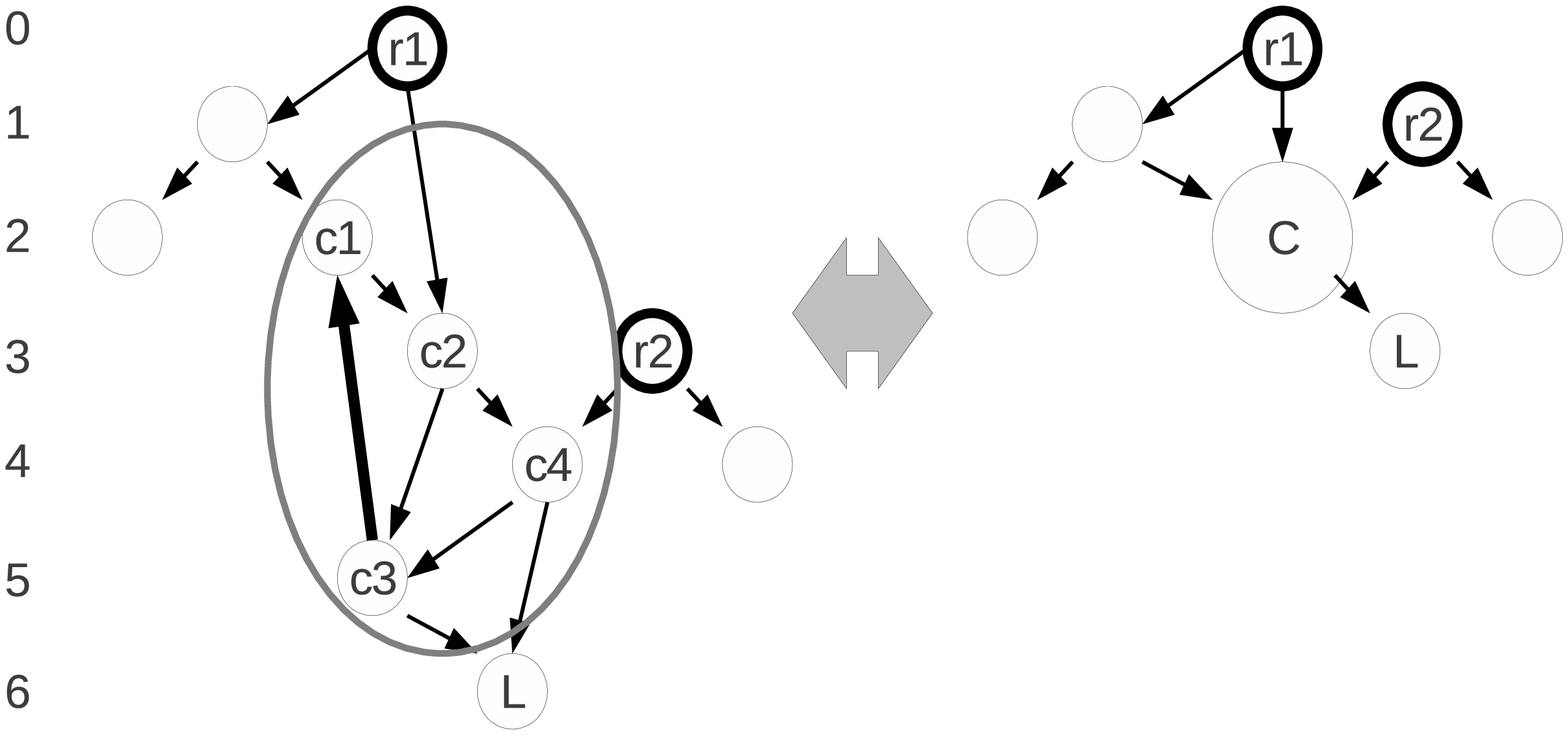}{The idea of loop-collapse of network -- virtual substitution of cycles with single ``complex'' vertices. Example network is the same as in Figure \ref{ksjh_fig_alternative} except single added arc between vertices $c3$ and $c1$ (depicted thicker). All vertices marked as $c1$,$c2$,$c3$ and $c4$ form two non-separated loops, and thus are collapsed into single vertex $C$. After finding out depth values of collapsed network, vertices $c1$,$c2$,$c3$ and $c4$ are assigned depth corresponding to vertex $C$. It is possible to have a loop as a ``root'' in the full network when using loop-collapse, if all vertices in that loop only have incoming arcs from other vertices in that loop. Note that the relative depth of vertex $L$ and whole network decreases significantly due to introduction of the single $c3-c1$ arc.}

Both definitions attempt to measure the vertical size of the hierarchy or a directed network. Despite aiming for the same, they significantly differ from each other on several points.//
%Despite both definitions trying to characterize the vertical (along the directed edges) size of the network, they give have several properties that are significantly different between them.\\
The \emph{information reduction} is present in both cases. Rooted depth uses only shortest path, which means that it effectively ignores any arcs that are not part of a shortest path, which essentially reduces the whole network to a spanning tree of shortest paths. The relative depth on the other hand effectively ignores arcs that are not part of a longest path. Relative depth definition essentially takes into account only arcs that are part of transitive reduction of a relation defined by arcs (which by definition of relative depth is transitive). All ``shortcuts'' are ignored, often including the shortest paths.\\
Both definitions are different in the aspect of \emph{locality}. Rooted depth is defined globally and does not describe the local situation consistently -- depth levels viewed from local perspective of single vertex may be completely not related to the graph structure. The only rule is that outgoing arcs of given vertex $i$ always connect to vertices with depth no larger than $d_i+1$ and incoming arcs are form vertices with depths no smaller than $d_i-1$. The relation between depth levels does not otherwise reflect arc structure, but the depth level of $i$ accurately describes how close the vertex is to the root. Moreover the root is always at known, fixed depth level $0$. On the other hand relative depth is defined locally and as such is relevant to local structure. Arcs always go from lesser to greater depth levels, although the actual values are effect of global topology. The depth values do not reflect how close to the eventual root (the ``top'') is the vertex. A vertex at depth $1000$ could be 
connected directly from the root. Moreover, the depth of a root is not fixed, and if several roots exist, may be found at any depth. Overall, the rooted depth describes the global position more accurately, but fails locally. The relative depth vice versa.\\
\emph{Vulnerability} of both definitions are also different. Rooted depth requires existence of root vertices and cannot work without them, while the definition of relative depth doesn't work in presence of directed loops. In both cases the loop collapse allows them to work, but nonexistence of root is rare for a network that would not be almost wholly loop-collapsed into single vertex (at least for random graphs). Both definitions of depth are volatile in regard to addition or removal of arcs. A single removal or addition could change depth levels of significant parts of the whole network. Adding new links can only decrease depth levels for rooted depth, and increase them for relative depth, and vice versa.\\
The rooted depth can take into account the \emph{multiple aspects} of depth. If multiple roots are present, then the depth levels relative to each root can be treated separately, not averaged into a single number, or alternatively averaged with certain weights depending on the root or its rooted component. If the depth is considered in the context of a hierarchy, this can differentiate between positions within different hierarchies. Relative depth cannot distinguish multiple aspects, as it has to be defined with a single number for each vertex. It is however \emph{consistent}, as that number is relatively well defined, without multiple aspects and issues tied to averaging.

Overall, both definitions have their strong and weak points. As they describe the complex topology by simple numbers, they reduce information in different ways, which yields different properties. The choice of depth measure will therefore depend on the aspects of the system being investigated and what one wishes to describe with it.

\section{Results}
\label{ksjh_sect_results}
We have investigated the depth measure values in directed random networks. We investigated a directed Erdos-Renyi graph, where probability of an arc existing between an ordered pair of vertices $ij$ is constant $p=\mean{k}/N$. Opposite arcs (e.g. $ij$ and $ji$) can exist and their existence is decided independently. Note that the parameter $\mean{k}$ is a directed degree -- $\mean{k}=k_0$ means that on average vertices have $k_0$ outgoing arcs as well as $k_0$ incoming arcs. If we looked at those links as undirected the total degree would be twice as high.\\

First, we attempt to investigate most general properties of the depth depending on the network parameters. The average depth $D$ is a very simple measure, allowing us to look across different network parameters. We are interested in the total depth of the network however, not necessarily the average value of $d$ over all vertices. Similar to physical depth, being distance from surface to bottom, not some average over a volume, we decide to measure $D$ as average od depth level over leaves in the network. For rooted depth and single-root network, this means
\begin{equation} D=\mean{d}_{leaves}=\frac{1}{N_L} \sum_i l_{ri} \label{ksjh_eq_depth1simple} \end{equation}
where $N_L$ is the number of leaves, the sum over $i$ is over all leaves and $l_{ri}$ is the path length from the root $r$ to leaf $i$. If there is more than one root, the depth level of vertices is average over all roots \emph{it has directed paths from}, which means
\begin{equation}D=\mean{d}_{leaves}=\frac{1}{N_L} \sum_i d_i =\frac{1}{N_L} \sum_i \frac{1}{N_{Ri}} \sum_r l_{ri} \label{ksjh_eq_depth1} \end{equation}
where $N_L$ is the number of leaves, the sum over $i$ is over all leaves, $N_{Ri}$ is number of roots that have path to leaf $i$, the sum over $r$ is over all roots that have path to leaf $i$ and $l_{ri}$ is the path length from the root $r$ to leaf $i$. The averaging od depth level for each vertex makes sense when we are investigating that single vertex. If we are looking at whole network however, it is more reasonable to look at pairs root-leaf and averaging over that. This would yield a definition
\begin{equation}D=\mean{d}_{leaves}=\frac{1}{N_{RL}} \sum_{r,i} l_{ri} \label{ksjh_eq_depth1pair} \end{equation}
where $N_{RL}$ is number of pairs root-leaf connected by directed path, the sum is over all pairs $r,i$ that are connected by a path and $l_{ri}$ is the lengh of that path. Equations \ref{ksjh_eq_depth1} and \ref{ksjh_eq_depth1pair} define slightly different values. In the first equation, if vertex is under small number of roots, the depth levels generated by them are contributing more to the total compared to a vertex under large number of roots, which averages over more different depths. If all vertices have directed paths from all roots, then both equations are equivalent as $1/N_{Ri}=const.$ and therefore it can be extracted from under the sum over $i$ and $N_L N_R=N_RL$ is all pairs have paths between them, thus recovering equation \ref{ksjh_eq_depth1pair}. As we are calculating a global property, we decided to use the definition described by Equation \ref{ksjh_eq_depth1pair} as more appropriate on the global level.\\

The dependence of average \emph{rooted depth} $D$ on average node degree $\mean{k}$ is shown at Figure \ref{ksjh_fig_erdos1}. Note that in case where no roots or leaves could be found, we assumed depth equal to 0 for practical reasons.

\rysunek{ksjh_fig_erdos1}{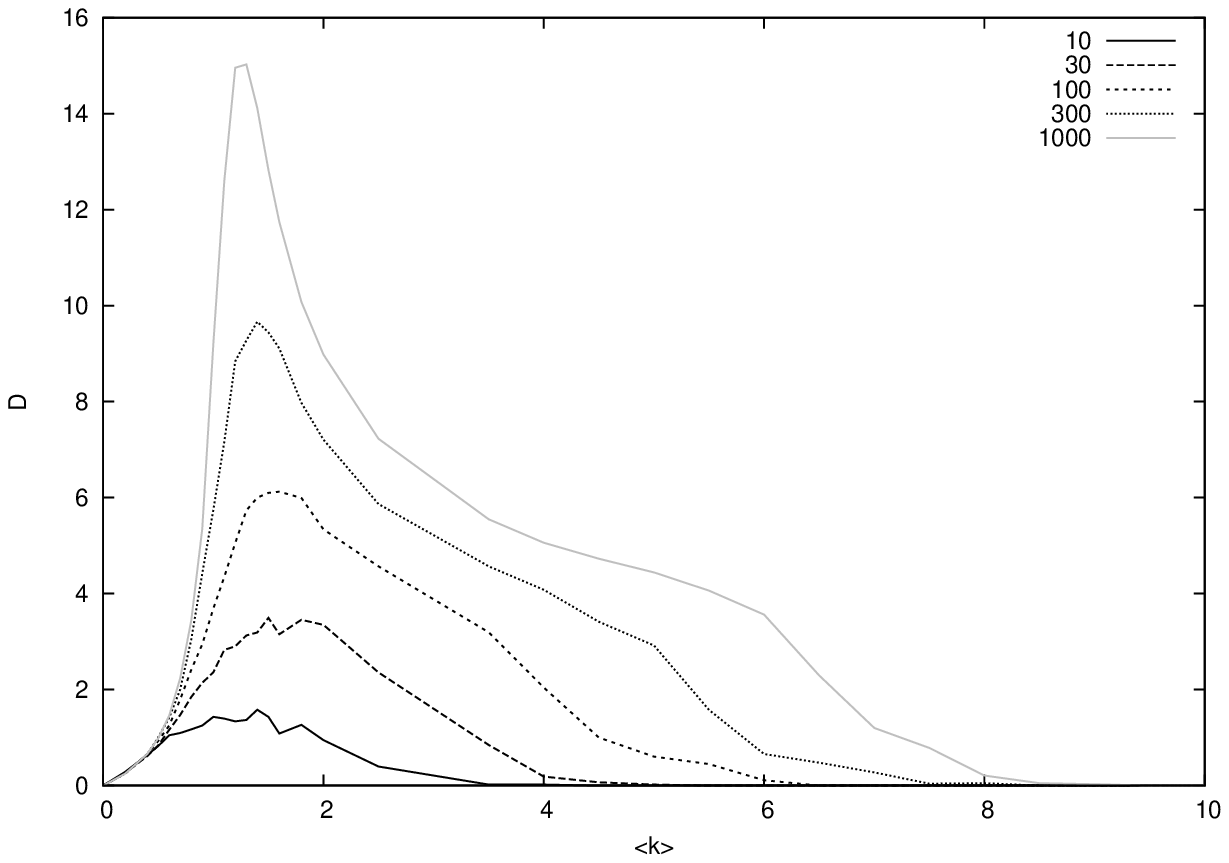}{Dependence of average rooted depth of directed ER graph on average directed degree for different network sizes.}

We explain the results at Figure \ref{ksjh_fig_erdos1} as follows. Below $\mean{k}=1$, the network is not percolated and consist of single vertices ($D=0$) and small clusters (responsible for increasing value of $D$ with $\mean{k}$). Loops are very rare. At $\mean{k}=1$ the network becomes percolated (note that this is directed percolation), creating the giant cluster. This corresponds with the maximum of the average depth level. The average depth is tied ot the diameter of the giant cluster as well as the amount of still present disconnected clusters. Above $\mean{k}=1$ adding more arcs means creating loops, since the network is already percolated. Loops introduced randomly cause decrease of the component diameter and thus depth, which corresponds to the first drop of $D$ after the peak. Increasing number of loops also may cause situations where no roots or no leaves exist in the network, causing the average depth to drop towards $0$ (the assumed value for network without roots or leaves).
This starts to happen only at high $\mean{k}$ and is responsible for second drop of the $H$ (around $\mean{k}=6$ for $N=1000$). Figure \ref{ksjh_fig_topology} shows how example network looks like at different $\mean{k}$.\\
It is notable that the shape of the dependence of network depth on density shown on Figure \ref{ksjh_fig_erdos1} is very similar to Global Reaching Centrality measure introduced by Mones et.al. \cite{ksjh_mones} and calculated for directed random graphs \cite{ksjh_mones2}. The similarity can be explained, by both depth and GRC being measures for hierarchies. Below percolation threshold, the graph consists of many small clusters, which yield very low average depth, at the same moment giving very low GRC due to narrow Local Reaching Centrality distributions. Significantly above percolation threshold, the presence of many loops mean that the network has large strongly connected component, that has relatively low depth as well as again producing narrow LRC distribution and in effect low GRC. At percolation threshold, the giant connected component is critically sparsely connected, meaning no loops and in effect a high depth (long shortest paths). On other hand, without loops the vertices have a broad distribution 
of LRC, resulting in high GRC. Both measures achieve high values for a large, extended directed network without loops, which explains the correlation.\\

\rysunek{ksjh_fig_topology}{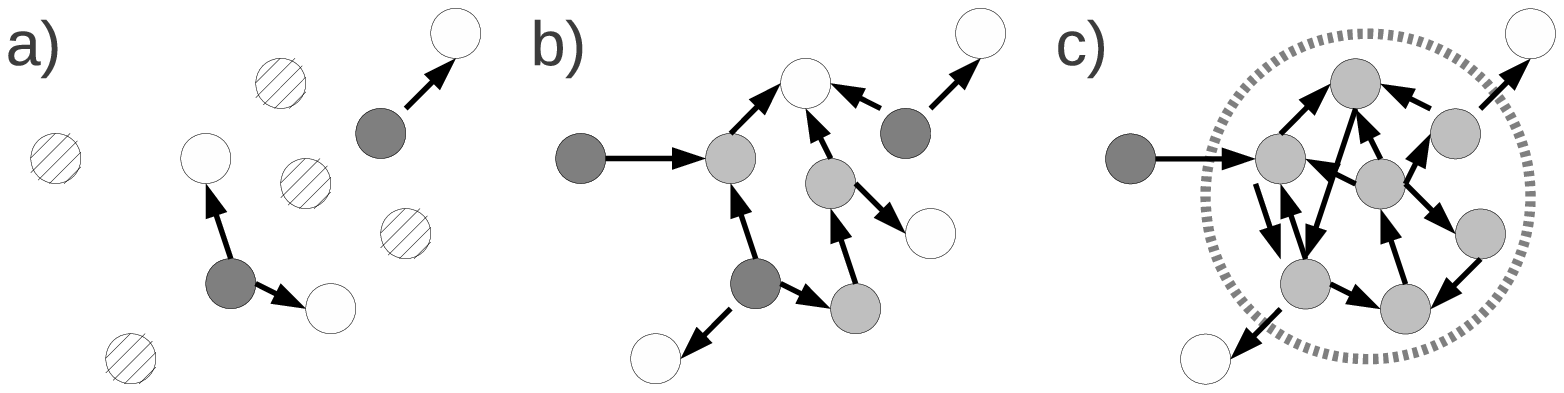}{The example topology of random network at different arc densities $\mean{k}$ and its relation to the depth values. \emph{a)} for low density $\mean{k}<1$, \emph{b)} at percolation threshold $\mean{k}=1$, \emph{c)} for high density $\mean{k}>1$. White vertices are leaves, dark gray are roots and hatched vertices are both leaves and roots (and in general single vertices). At \emph{c)}, the network has a strongly connected ``core'' that consists of interleaving loops and therefore under loop-collapse rules all have same depth. The roots and leaves are attached to that core.}

\rysunek{ksjh_fig_erdos2}{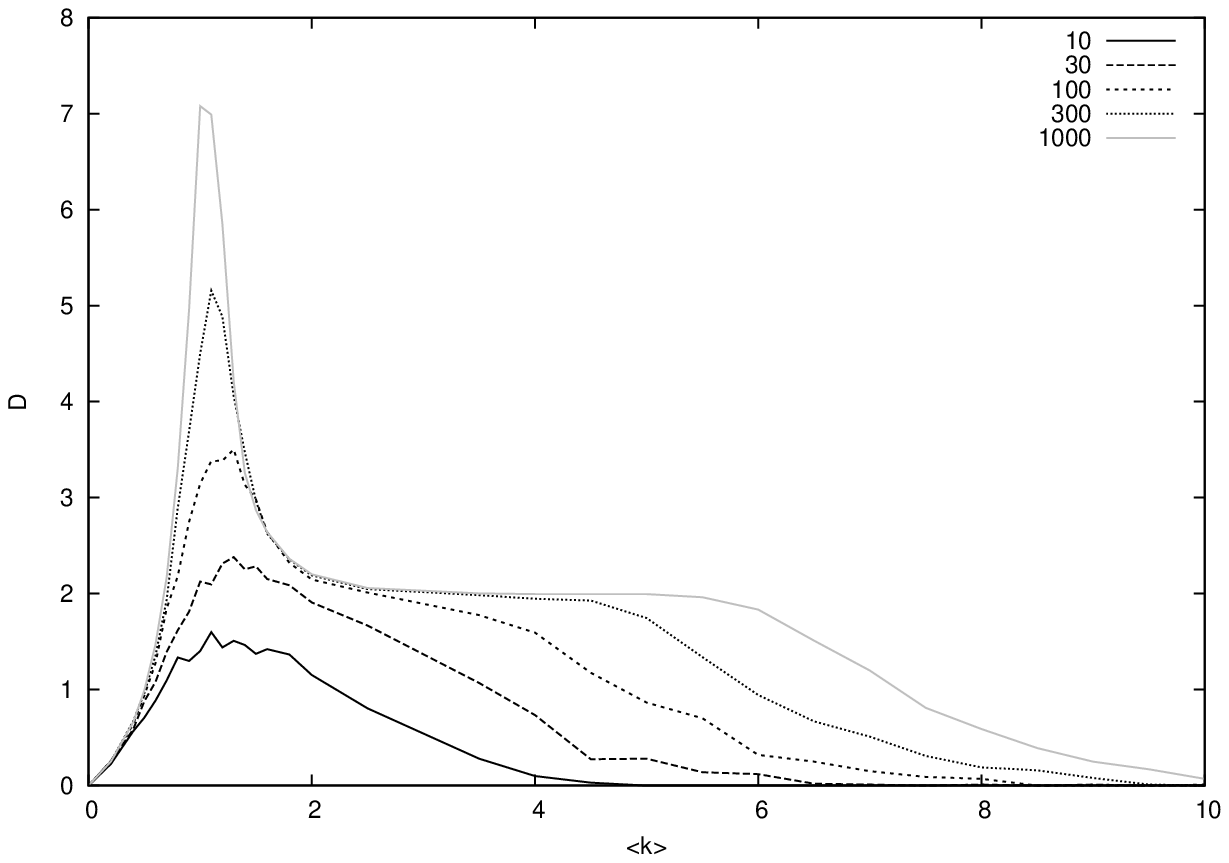}{Dependence of average rooted depth of loop-collapsed directed ER graph on average degree for different network sizes. Note that all the values are for the real, full network, not loop-collapsed one.}

Because the rooted depth cannot be resolved in case there are no roots or leaves in the network, and also to have better comparison between rooted depth and further relative depth, we decided to also investigate rooted depth while using the loop-collapse. The results are somewhat different and found at Figure \ref{ksjh_fig_erdos2}. The main differenc is the plateau after the peak, with value $D=2$. This corresponds to a network with a dense ``core'' containing many interleaving loops, and thus collapsed to a single complex vertex, with single vertex ``roots'' and ``leaves'' sticking out of this core. Without the loop-collapse, the loops were partially converted to ``vertical'' and assigned some differing depths (differently from perspective of each root), the denser the core, the smaller depths. With loop-collapse, the value of $D$ for network with the dense core is fixed at $2$, resulting in plateau.
Note that since loop-collapse virtually removes all loops, there are always roots and leaves and in worst case the sole remaining complex vertex is both root and leaf and thus has $D=0$. Also note that all the values ($D$, $\mean{k}$, $N$) are calculated for the real, full network. Loop-collapse is used only to calculate depth levels.\\

The results are similar, although there are differences. Below $\mean{k}=1$, the behavior is the same, as lack of loops means loop-collapse doesn't do anything. After the maximum, the average depth $D$ declines much faster, because addition of loops causes loop-collapse, thus decreasing the effective size of the giant component much quicker. The value then stabilizes at $D=2$, because the typical topology of loop-collapsed network at this point consists of single complex network representing the majority of original vertices, with few single roots and leaves directly attached to it. As the $\mean{k}$ becomes even higher, the network starts to entirely collapse to single complex network, causing $D$ to approach $0$ for very high $\mean{k}$. This explanation is reinforced by the size of the loop-collapsed network, that is very close to original size below percolation threshold $\mean{k}=1$, and then exponentially decreases towards single vertex (Figure \ref{ksjh_fig_size1}).

\rysunek{ksjh_fig_size1}{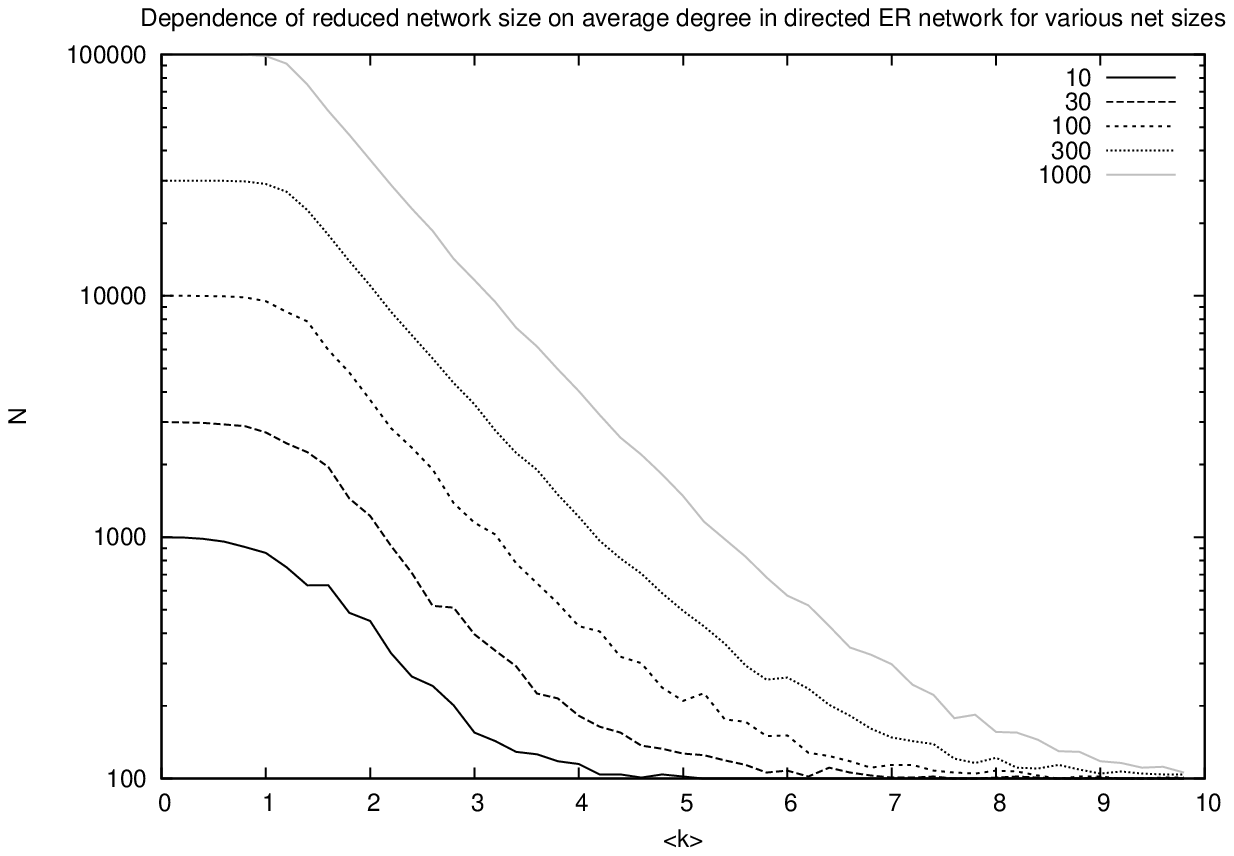}{Dependence of size of loop-collapsed directed ER graph on original average (in/out, not total) degree for different original network sizes.}

The measurement of \emph{relative depth} can be performed similarly. Since we need to use loop-collapse, we always recover roots and leaves. If our relative depth levels are scaled so that the lowest value is $0$ (similarly to how roots in rooted depth have depth $0$), we can calculate average depth of leaves (the ``bottom'') and treat that as the depth of the network
\begin{equation} D=\mean{d}_{leaves} \label{ksjh_eq_depth2} \end{equation}
where $\mean{d}$ is average over leaves. Note however, that unlike in rooted depth, roots (the ``surface'') can be at different depth levels themselves. We can thus calculate the depth of whole network as difference between average depth of leaves and roots (the ``bottom'' and ``surface'')
\begin{equation} D=\mean{d}_{leaves}-\mean{d}_{roots} \label{ksjh_eq_depth2r} \end{equation}.
The first equation (\ref{ksjh_eq_depth2r}) describes the depth of the network globally, taking into account the extreme value (for roots). The second equation (\ref{ksjh_eq_depth2}) is a more locally relevant, showing what is the likely depth in a part of the network. It could be compared to rooted depth where we only take ``highest'' root into account and normal rooted depth that considers all roots. Both equations yield similar results, presented in Figure \ref{ksjh_fig_erdos3}. In general, it can be seen that relative depth behaves similarly to rooted depth. For small $\mean{k}$ the value is low. The peak is around percolation threshold, corresponding to the maximally stretched treelike structure of the giant component, and then the decline of average depth correspond to the gradual increase in density and number of loops, which flattens the network when if comes to depth (remember we use loop-collapse).
The one difference between rooted and relative depth discrenible on the plots is the absence of the sudden drop for higher $\mean{k}\approx 6$. This can be explained by the fact, that when we couldn't find either roots or leaves in rooted depth calculation, we assumed total depth equal to $0$. If we look at Figure \ref{ksjh_fig_topology}, we see that in the ``core with roots and leaves'' stadium, loss of last root effectively makes the value $0$ (the assumed value), despite one arc earlier having depth $2$ or even more. This accounts for the fast drop of depth for rooted depth not using loop-collapse. Since relative depth always uses loop collapse, the ``core with roots and leaves'' would simply change into depth $1$ structure, where either the ``core'' would assume role of leaf/leaves (loss of last leaf) or role of root (loss of last root). The transition is smooth, thus no sudden drops on the graphs for relative depth could be found.

%!!!!!!!!!!!!!!!!Tutaj dalej bede musial jeszcze poprawic, wlacznie z wykresem dla ``relative depth'' dla pikow
%The peak is higher, because in the giant cluster just above percolation threshold, large directed loops are still rare, thus not evoking loop-collapse and same depth levels on large number of vertices, but the longest path relevant to the definition is significantly larger than the shortest path used in rooted depth. 

\rysunek{ksjh_fig_erdos3}{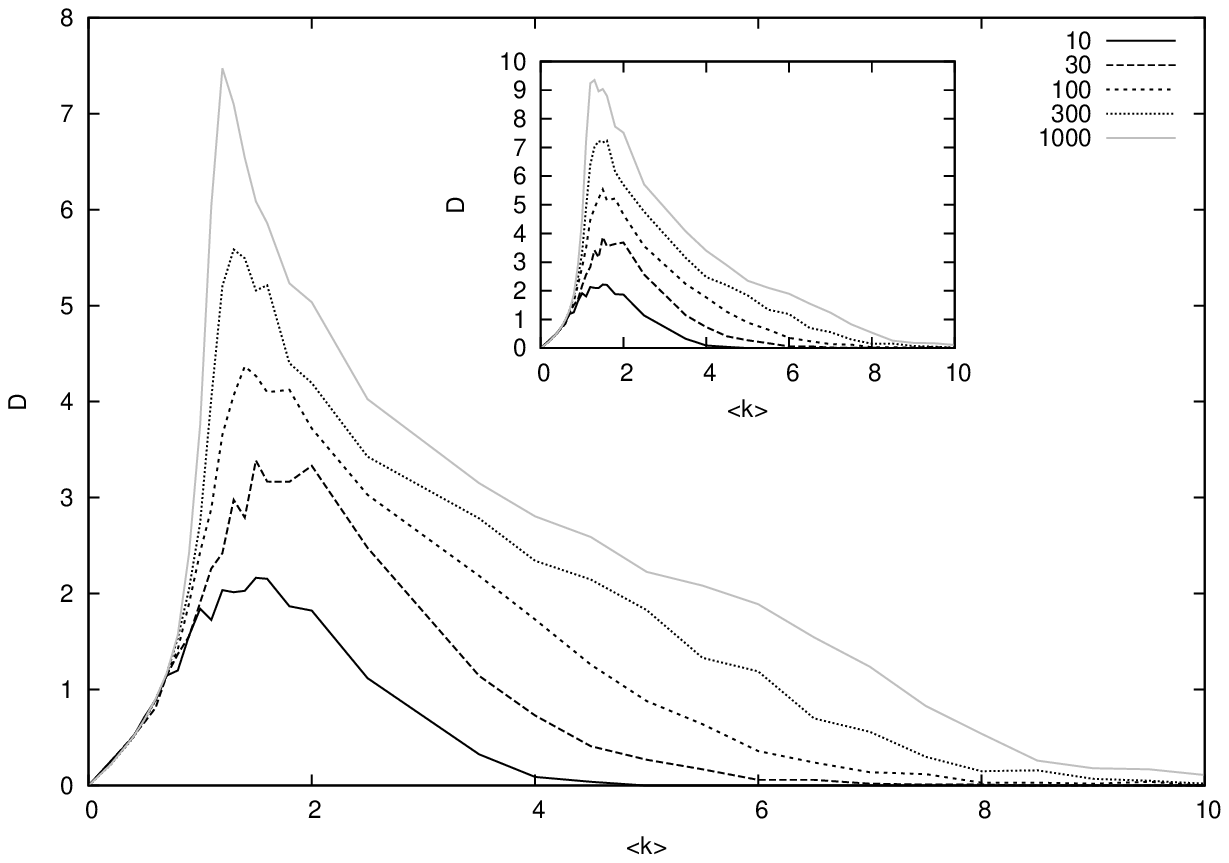}{Dependence of average relative depth of loop-collapsed directed ER graph on average degree for different network sizes, computed as difference between leaf and root average depths (Rq.\ref{ksjh_eq_depth2r}). \emph{Inset:} Plot for relative depth scaled to $0$ for lowest-depth root. Both plots are essentially the same, with small differences in heights.}

%\rysunek{ksjh_fig_erdos3}{erdos_new_cd.eps}{Dependence of average relative depth of loop-collapsed directed ER graph on average degree for different network sizes, computed as average over leaves (Eq.\ref{ksjh_eq_depth2}). Note that the relative depth is scaled as to equal $0$ for lowest-depth roots.}

%\rysunek{ksjh_fig_erdos3r}{erdos_new_cdr.eps}{Dependence of average relative depth of loop-collapsed directed ER graph on average degree for different network sizes, computed as difference between mean depth of leaves and roots (Eq.\ref{ksjh_eq_depth2r}).}

Trying to describe the behavior of the depth measures for different network parameters we also investigated the height of the $D$ peak for different network sizes and different definitions. Figure \ref{ksjh_fig_peaks} shows results for rooted depth of networks without loop-collapse, with loop-collapse and for relative depth. We found power dependence od rooted depth on size, with exponent $\approx 0.3$ to $\approx 0.4$). We expect these values to be related to the diameter (maximum shortest path length) of the percolation cluster in random graph. The relative depth peaks also behave as a power of network size, albeit with significantly lower exponent $\approx 0.2$.

\rysunek{ksjh_fig_peaks}{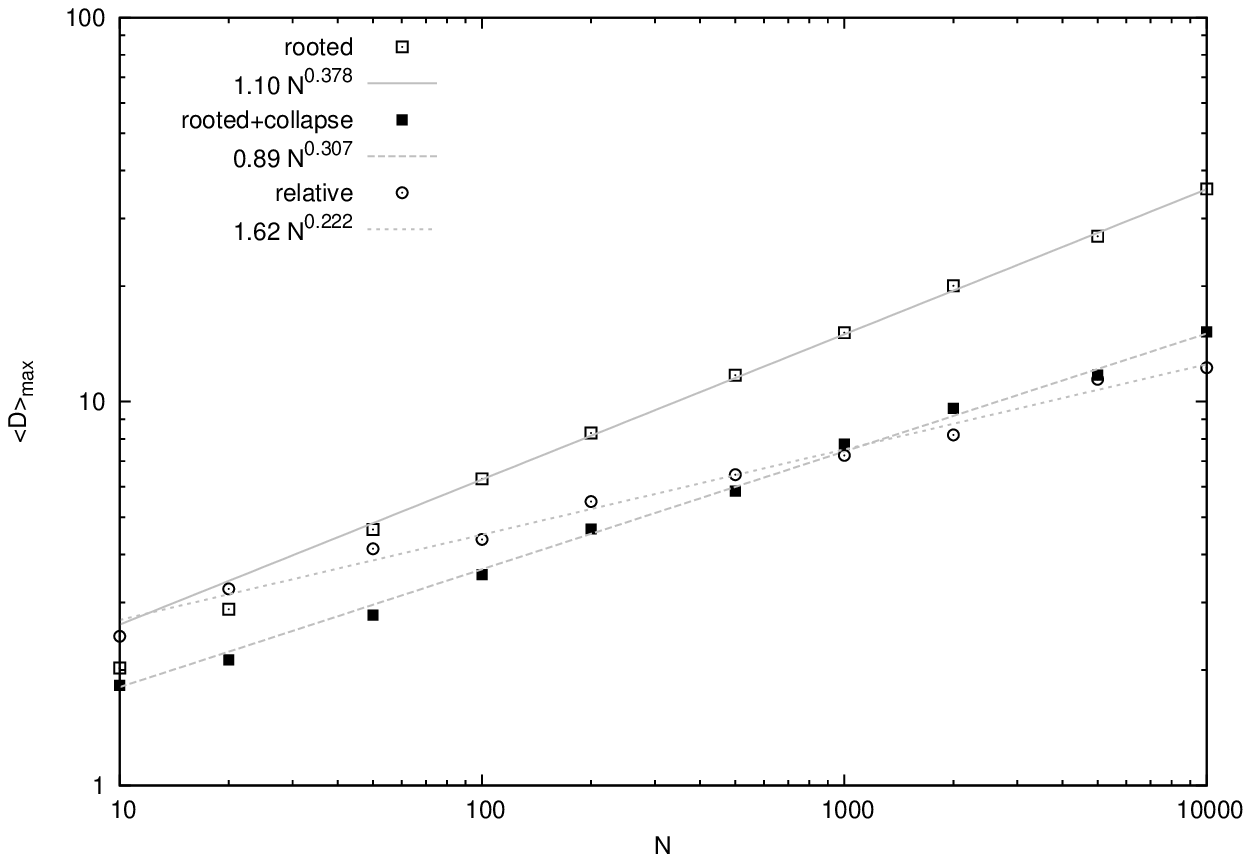}{Dependence depth peak heights for ER graphs on the graph size. The three series are rooted depth without loop-collapse, with loop-collapse and relative depth peak heights. The height of the peak was measured from a series of $D(\mean{k})$ with $0.1$ resolution on the $\mean{k}$ axis.}
%\rysunek{ksjh_fig_peaksnc}{erdos_peaks_nc.eps}{Dependence of rooted depth peak height for ER graphs on the graph size.}
%\rysunek{ksjh_fig_peaksc}{erdos_peaks_c.eps}{Dependence of rooted depth peak height for loop-collapsed ER graphs on the original graph size.}
%\rysunek{ksjh_fig_peakscd}{erdos_peaks_cd.eps}{Dependence of relative depth peak height for loop-collapsed ER graphs on the original graph size.}

\section{Conclusions}
We have defined two depth measures for flow hierarchies, generalized to any type of directed network. \emph{Rooted depth} is defined as shortest path from one of network's roots, while \emph{relative depth} is defined through relations between vertices and is effectively equal the longest path from the root. The differences between these measures are discussed -- rooted depth ignores arcs that are not shortest paths, while relative depth ignores arcs that are not longest paths. Rooted depth has more global meaning, while relative one is more meaningful locally. We have investigated the two measures on a random Erdos-Renyi networks, showing how they behave on purely random network, thus allowing better understanding of values obtained in other types of networks. Both measures behave similarly to a Global Reaching Centrality measure of how hierarchical the network is.\\

\section*{Acknowledgements}
The research leading to these results has received funding from the European Union Seventh Framework Programme (FP7/2007-2013) under grant agreement no 317534 (the Sophocles project).

\end{document}